\documentclass[letter,tradiabstract]{aa}  
\pdfoutput=1

\usepackage{graphicx}
\usepackage[caption=false,position=top,font=scriptsize,labelformat=empty,aboveskip=0pt,captionskip=0pt]{subfig}
\usepackage[draft]{hyperref}
\usepackage{amssymb}
\usepackage{amsmath}
\usepackage{lmodern}
\usepackage{textcomp}
\usepackage{float}
\usepackage{threeparttable}
\usepackage{tikz}
\usetikzlibrary{arrows,shapes,backgrounds}
\usepackage{natbib}
\bibpunct{(}{)}{;}{a}{}{,}
\usepackage{txfonts}
\begin{document} 
\title{Large dust grains in the wind of VY Canis Majoris\thanks{Based on observations made with European Southern Observatory (ESO) telescopes at the La Silla Paranal Observatory under program 60.A-9368(A).}}

\author
{P. Scicluna\inst{1} \and R. Siebenmorgen\inst{2} \and R. Wesson\inst{3} \and J.A.D.L Blommaert\inst{4} \and M. Kasper\inst{2} \and N.V. Voshchinnikov\inst{5} \and S. Wolf\inst{1} }
\institute{
ITAP, Universit\"at zu Kiel, Leibnizstr. 15, 24118 Kiel, Germany 
 \and
European Southern Observatory, Karl-Schwarzschild-Str. 2, D-85748 Garching b. M\"unchen, Germany
 \and
European Southern Observatory, Alonso de C\'{o}rdova 3107, Casilla 19001, Santiago, Chile
 \and
Astronomy and Astrophysics Research Group, Dep. of Physics and Astrophysics, V.U. Brussel, Pleinlaan 2, 1050 Brussels, Belgium
 \and
Sobolev Astronomical Institute, St. Petersburg University, Universitetskii prosp., 28, St. Petersburg 198504, Russia
}

\date{}

\abstract{
Massive stars live short lives, losing large amounts of mass through their stellar wind. 
Their mass is a key factor determining how and when they explode as supernovae, enriching the interstellar medium with heavy elements and dust. 
During the red supergiant phase, mass-loss rates increase prodigiously, but the driving mechanism has proven elusive. 
Here we present high-contrast optical polarimetric-imaging observations of the extreme red supergiant VY Canis Majoris and its clumpy, dusty, mass-loss envelope, 
using the new extreme-adaptive-optics instrument SPHERE at the VLT. 
These observations allow us to make the first direct and unambiguous detection of submicron dust grains in the ejecta; we derive an average grain radius $\sim$\,0.5\,$\mu$m, 50 times larger than in the diffuse ISM, large enough to receive significant radiation pressure by photon scattering. 
We find evidence for varying grain sizes throughout the ejecta, highlighting the dynamical nature of the envelope. 
Grains with 0.5\,$\mu$m sizes are likely to reach a safe distance from the eventual explosion of VY Canis Majoris; hence it may inject upwards of 10$^{-2}$\,M$_\odot$ of dust into the ISM. 
}

\maketitle

\section{Introduction}

When massive stars (M$_\ast \geq 8$\,M$_\odot$) approach the end of their lives, they may expand and cool to become red supergiants (RSGs), a phase characterised by prodigious mass loss and dust production.
The source of momentum for the wind remains unclear: the acceleration of dust grains by radiation pressure seems the most promising mechanism, but the formation of grains is inhibited by the extreme luminosity of RSGs.
The outer atmospheres of RSGs are $\sim$\,20\% larger than expected for a stellar photosphere and have high molecular abundances - this extension is likely to be caused by radiation pressure on molecular lines \citep{2015A&A...575A..50A}, but the dust species that can form at these distances ($< 3$\,R$_\ast$) are not abundant enough to supply the momentum required to drive the wind \citep{2012A&A...546A..76B}, while the iron-rich silicates that most efficiently absorb the star's radiation cannot form until $\sim$\,20\,R$_\ast$ \citep[extrapolating from the study of AGB stars in][]{2006A&A...460L...9W}. 
However, the acceleration of the wind to terminal velocity is observed to begin around 10\,R$_\ast$ \citep{1998MNRAS.299..319R}; 
the only abundant species that can form so close are Ca-Al or Mg rich silicates (e.g. melilite, Ca$_2$Al$_2$SiO$_7$, forsterite, Mg$_2$SiO$_4$ \citealt{2000A&AS..146..437S}), which are nearly transparent to the stellar radiation. 
If the acceleration is dominated by these species, they can only drive the wind if they are large enough to be highly efficient scatterers \citep{2008A&A...491L...1H,2012A&A...546A..76B}.

VY Canis Majoris (VY~CMa) is a nearby (1.2\,kpc, \citealt{2012ApJ...744...23Z}) dust-enshrouded RSG, a rare class of objects of which it is the best studied example.
It is one of the most intrinsically luminous stars in the sky ($L=3\times 10^5 L_\odot$, \citealt{2012A&A...540L..12W}) and, excluding solar-system bodies, is the third brightest object in the sky at 10\,$\mu$m \citep{1994yCat.2125....0J}.
It boasts a large optical reflection nebula, visible through even small telescopes, created by its prodigious mass loss ($> 10^{-4} {\rm M_\odot\,yr^{-1}}$, \citealt{2015A&A...573L...1O}).
Despite decades of detailed study \citep[e.g.][]{1969ApJ...156L.139S,1974ApJ...188..533H,1998AJ....115.1592K}, it remains enigmatic, and different wavelength regimes frequently resulting in conflicting results \citep{2009AJ....137.3558S,2015A&A...573L...1O}.
Its incredible brightness makes VY~CMa key to revealing the properties and evolution of dust in the mass-loss envelopes of RSGs.

Here, we present optical polarimetric imaging of VY~CMa obtained with the extreme adapative optics imager SPHERE (Sect.~\ref{sec_obs}), which resolves the ejecta in unprecedented detail.
Based on these data, we determine grain sizes for a number of features in the ejecta (Sect.~\ref{sec_res}). 
Finally, in Sect.~\ref{sec_dis} we discuss the implications of our findings, with particular regard for the need to understand wind driving and the fate of grains in a supernova explosion.

\section{Observations}\label{sec_obs}
We observed VY~CMa using the new Spectro-Polarimetric High-contrast Exoplanet REsearch (SPHERE) instrument at the Very Large Telescope (VLT) \citep{2006Msngr.125...29B}. 
This instrument combines extreme adaptive optics (XAO) with optical and near-infrared imaging to achieve very high contrast ($>10^{6}$). 
The sub-instrument, the Zurich IMaging POLarimeter (ZIMPOL) allows, for the first time, diffraction-limited observations in the visual with a telescope with an 8m mirror \citep{2008SPIE.7014E..3FT}.
VY~CMa was observed on the night of 8 December 2014 in V-band ($\lambda_{\rm eff} =$ 554\,nm) and narrow I-band ($\lambda_{\rm eff} =$ 816.8\,nm) filters using a Classical Lyot Coronagraph (CLC) with diameter 155\,mas to suppress the stellar point spread function (PSF).
Both detectors were exposed in the same filter simultaneously, so that the exposure time could be adjusted for optimal exposure in both filters.
Because VY~CMa is very bright ($m_V \sim 7$), observations were taken with fast polarisation modulation ($\sim$1\,kHz) to minimise atmospheric effects at the expense of sensitivity. 
The rotation of the instrument was compensated for using the active field derotation mode, in which an additional half-wave plate is used to ensure that the polarisation orientation remains constant, at the expense of increased instrumental polarisation ($\sim 0.5\%$ rather than $\sim 0.1\%$); this was deemed acceptable as previous polarimetric observations \citep{2007AJ....133.2730J} of VY~CMa have shown extremely high polarisation fractions such that any instrumental signal is negligible.

\begin{figure*} 
\subfloat[]{\includegraphics[scale=0.5,clip=true,trim=1.cm 1.cm 1.5cm 1.3cm]{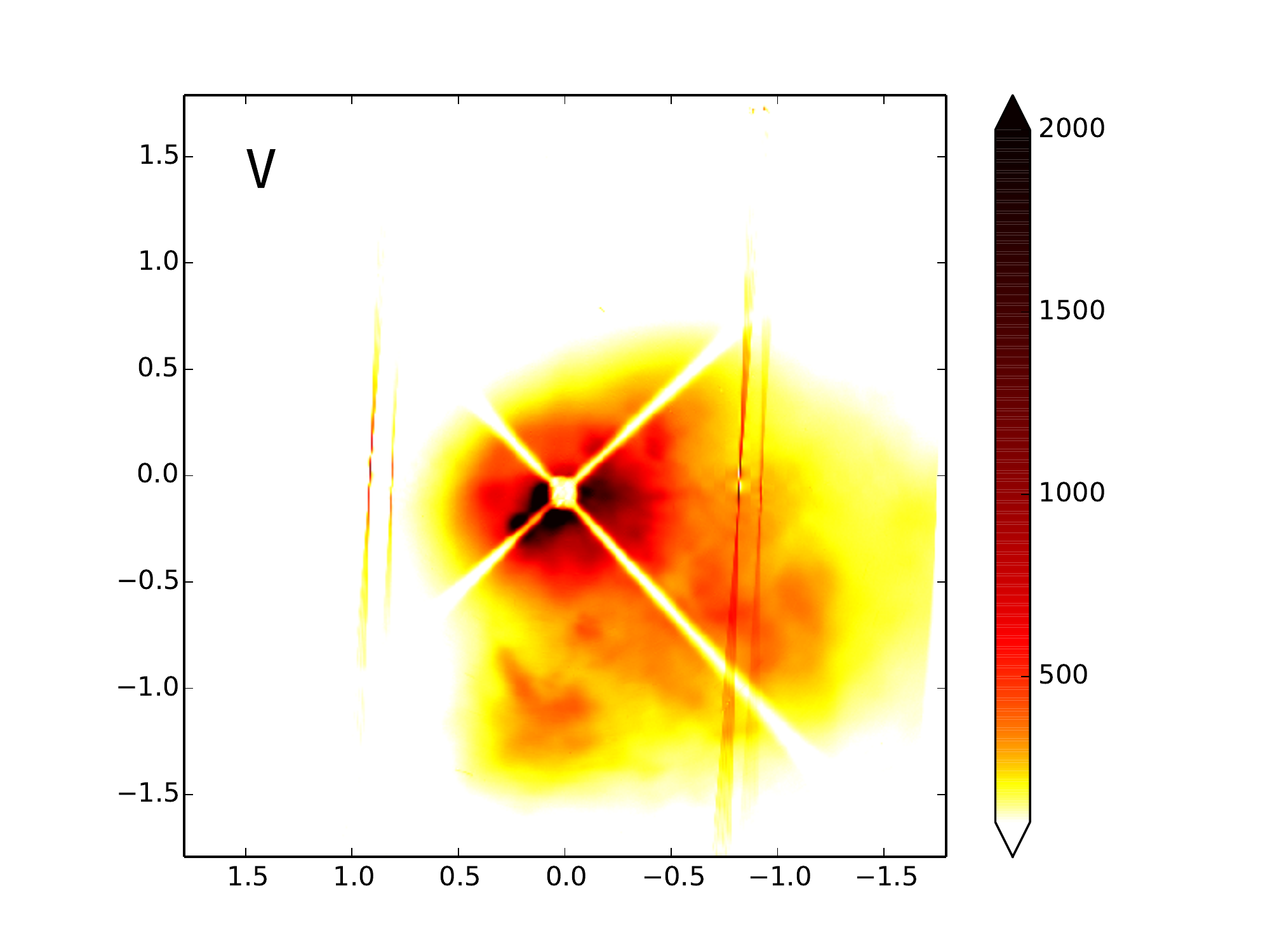}}	
\subfloat[]{\includegraphics[scale=0.5,clip=true,trim=1.cm 1.cm 1.5cm 1.3cm]{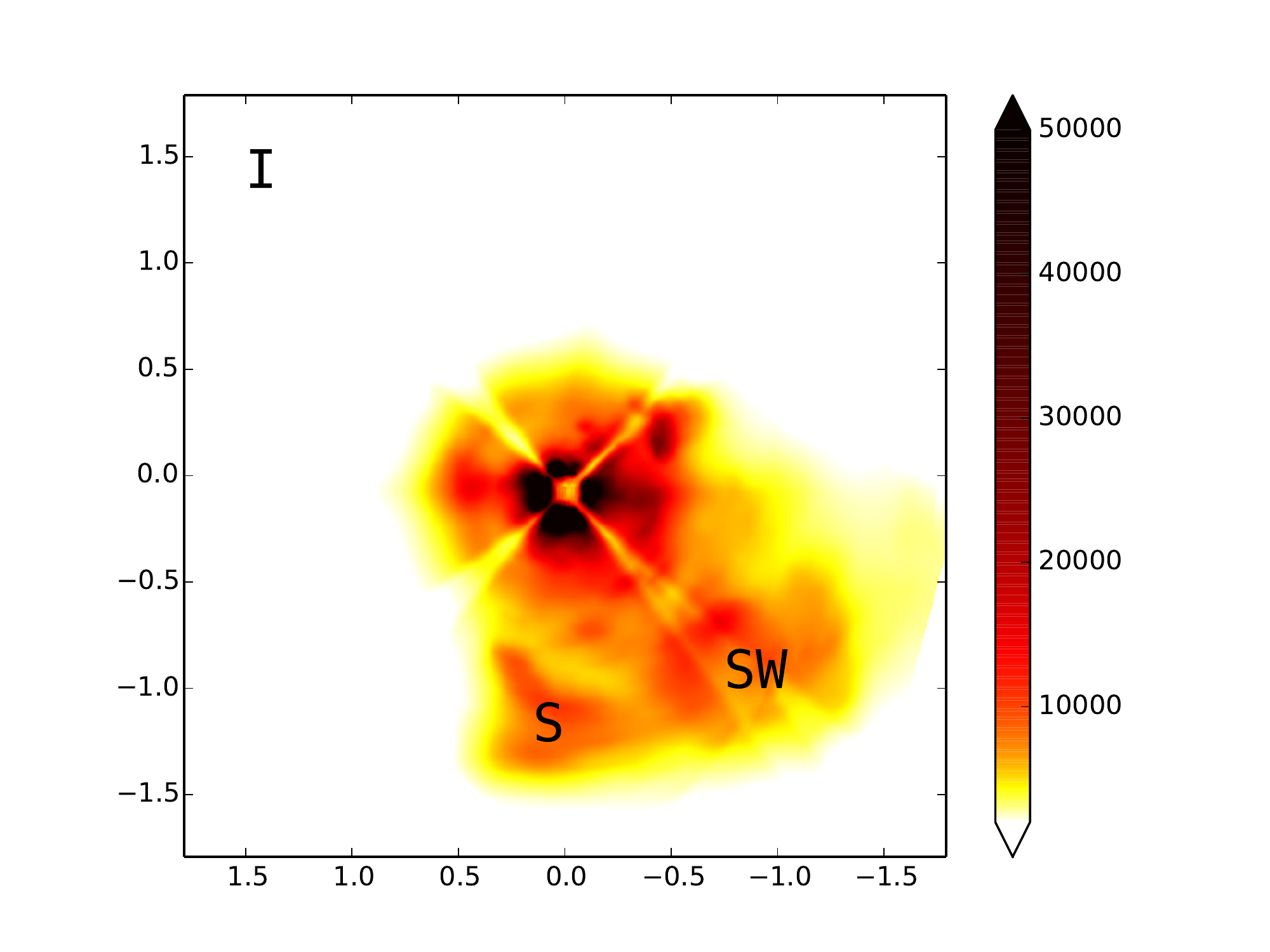}}	

\subfloat[]{\includegraphics[scale=0.5,clip=true,trim=1.cm 1.cm 1.5cm 1.3cm]{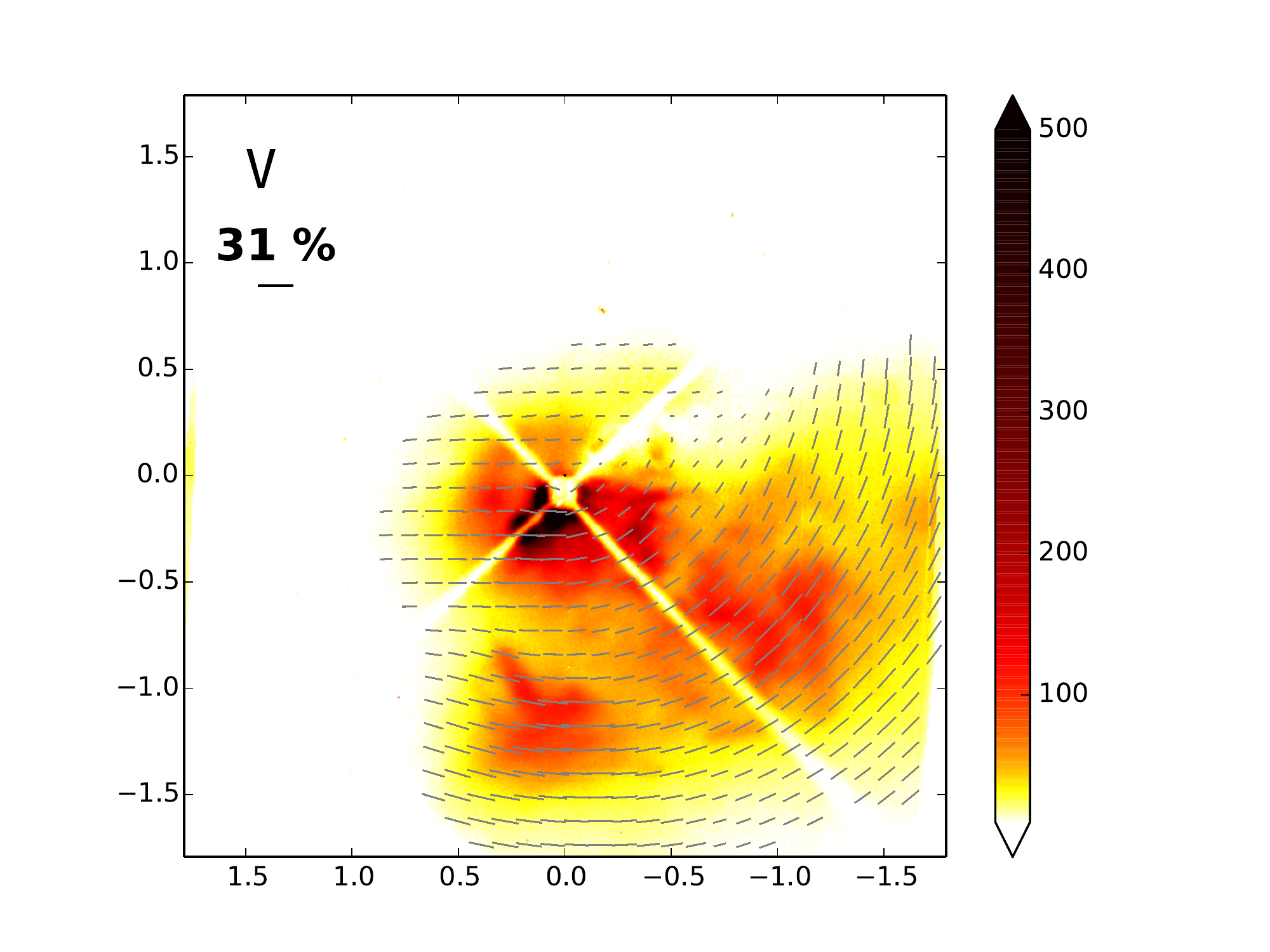}}	
\subfloat[]{\includegraphics[scale=0.5,clip=true,trim=1.cm 1.cm 1.5cm 1.3cm]{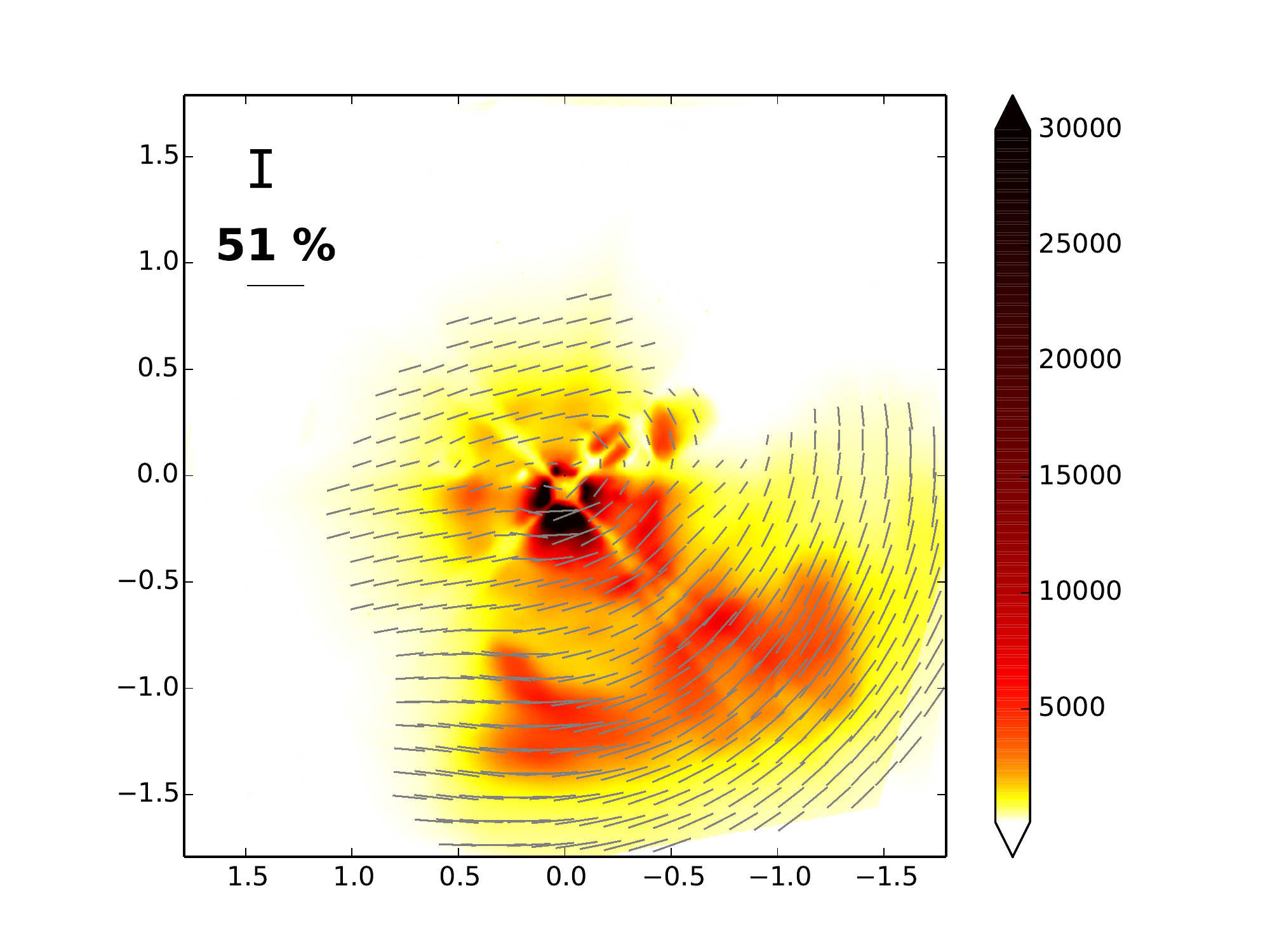}}	
\caption{{ Coronagraphic imaging polarimetry of VY~CMa.} {\it Top:} Intensity observations. {\it Bottom:} Polarised intensity, with polarisation direction and fraction shown in the overlaid vector field (only shown where both the fractional polarisation and the polarised intensity are significant). Offsets from the central star are shown in arcseconds, the colour scale in arbitrary units and the observing filter in the top left corner. The contrast is noticeably higher in polarised intensity, because the stellar halo is only weakly polarised. This allows a clearer identification of circumstellar material, and separation into discrete structures. The positions of the South Knot (S) and Southwest Clump (SW) are indicated in the I-band intensity image. One arcsecond corresponds to a projected distance of approximately 200 stellar radii. The low-intensity cross is produced by the suspended coronagraphic mask, which obscures part of the image plane, while the stripe-like artefacts in the V-band image result from a detector bias effect that cancels out in the polarised signal.}
\label{fig_zimpol}
\end{figure*}

The resulting data were reduced using the v\,0.14 of the {SPHERE} pipeline. 
Bias and dark--current subtraction was performed, followed by intensity flat--fielding. 
The pipeline then reconstructs the positive and negative Q and U signals for each detector, along with the Stokes I signal for each observation.
Then, an overall Q signal is derived from each detector using $Q=0.5\left(Q_{+} - Q_{-}\right)$, and similarly for U.
The resulting I, Q and U images from each detector were then co-added in python, before determining images of the polarised intensity, polarisation fraction and polarisation angle.

Figure \ref{fig_zimpol} shows the images obtained in both polarised and unpolarised light.
The polarisation images show a clear centro-symmetric pattern, indicating that the polarisation is caused by the scattering of light by dust grains.
As a result, we can use the fractional polarisation and the intensity ratio to constrain the properties of the scattering dust, in particular the size distribution of the grains.
The very high maximum polarisation degree ($\sim50\%$) indicates the presence of grains with radii similar to the wavelength of the observations ($\geq$100\,nm), so we calculate the maximum likelihood values of the minimum (a$_\mathrm{min}$) and maximum (a$_\mathrm{max}$) radii for a grain size distribution, assuming oxygen-deficient silicates \citep{1992A&A...261..567O}. 
To improve the constraints, we also incorporate the published H$\alpha$--polarisation measurements of \citet{2007AJ....133.2730J}.
Further details are given in the online appendix. 

\section{Results}\label{sec_res}

We considered several regions within the ejecta separately. 
Previous studies \citep{2007AJ....133.2716H,2007AJ....133.2730J,2007ApJ...656.1109M,2009AJ....137.3558S} have shown that VY~CMa's mass-loss envelope consists of a number of fast-moving, randomly directed ejection events (``clumps'') and slower, steady, equatorially enhanced mass-loss (the ``disc''); these studies have determined the three-dimensional geometry of the envelope, which we exploit to determine grain sizes.
The clumps are seen as distinct features in both intensity and polarisation in the ZIMPOL data. 

\begin{figure}
    	\resizebox{\hsize}{!}{\includegraphics{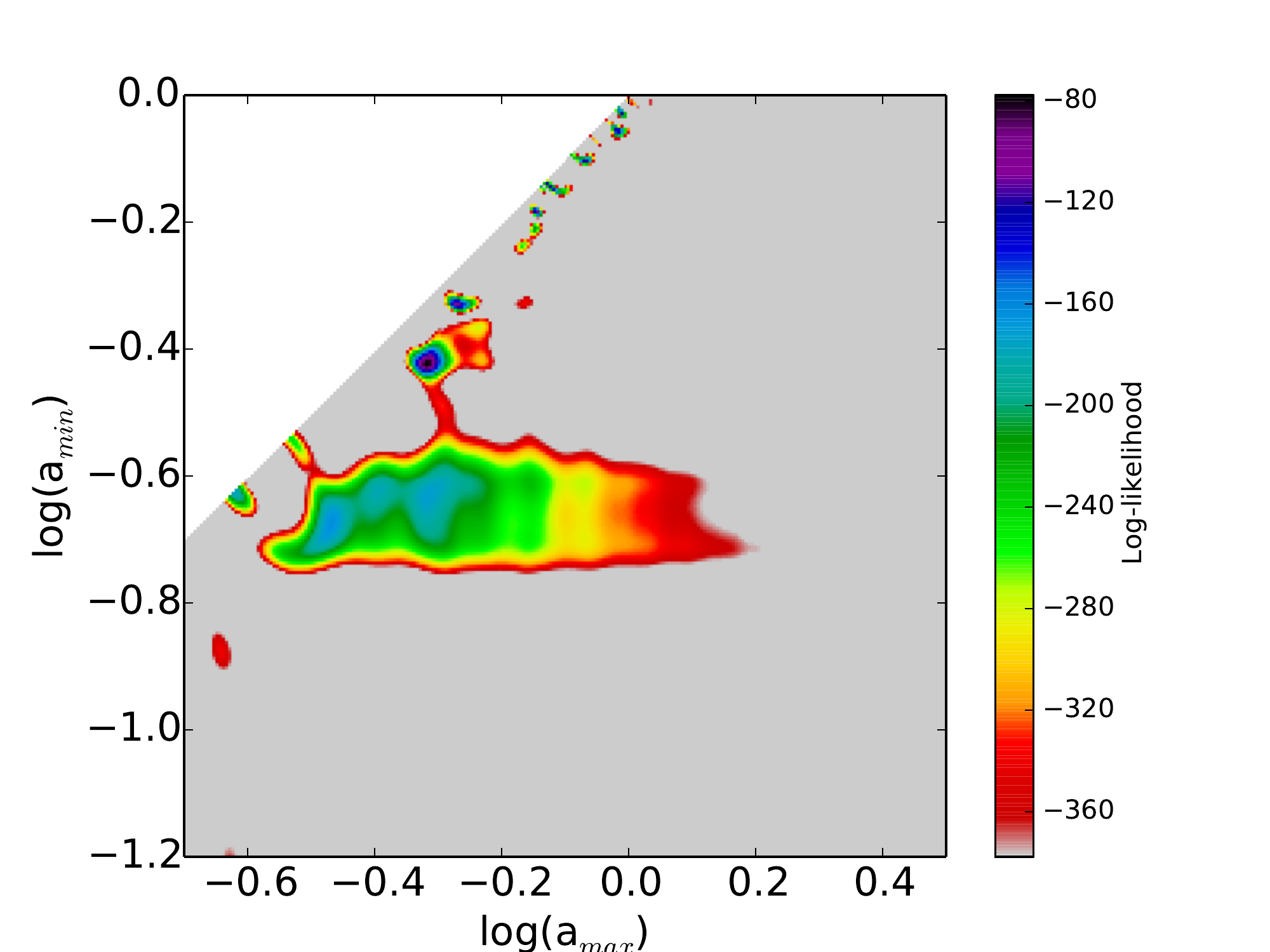}}
\caption{Likelihood space for a scattering angle of 98$^\circ$, corresponding to the SW Clump, zoomed to show grain sizes in the range 0.1\,--\,3\,$\mu$m. The equivalent figure for the S Knot is shown in the online appendix.}
\label{fig_like}
\end{figure}

We examine two clumps, the South Knot and the Southwest Clump (see Fig.~\ref{fig_zimpol}).
The angle joining the star, the South Knot and the observer is 117$^\circ$ \citep{2007AJ....133.2716H}, giving a scattering angle of 63$^\circ$. 
Log-likelihoods were calculated for a grid of grain-size distributions to find maximum likelihood values of a$_{\rm min}$ and a$_{\rm max}$ and explore a relevant parameter space (see Sect.~\ref{sec_obs} and Appendix~\ref{sec_fit}).
The parameter space consists of a number of islands in the region a$_{\rm min} \geq 0.25$\,$\mu$m that produce good fits with a$_{\rm max}$ only slightly larger than a$_{\rm min}$, i.e. approximately monodisperse. 
The best solution lies at a$_{\rm min}$ = 0.55\,$\mu$m and a$_{\rm max}$ = 0.58\,$\mu$m, with an average size of 0.56\,$\mu$m. 

The Southwest Clump, with a scattering angle of 98$^\circ$ \citep{2007AJ....133.2716H}, shows similar behaviour (see Fig.~\ref{fig_like}), but with the maximum likelihood model (a$_\mathrm{min}=$\,0.38\,$\mu$m, a$_\mathrm{max}=$\,0.48\,$\mu$m) requiring a broader range of grain sizes. 
This implies an average size of 0.42\,$\mu$m.
However, as these regions may be optically thick, multiple scattering may play a role, reducing the observed polarisation fraction in these regions. 
As a result, while this indicates differences between the quiescent and clump phases, polarimetric observations in the near-infrared -- where the clumps are optically thinner -- would confirm our findings. 

\section{Discussion}\label{sec_dis}

\subsection{Large grains}

Grains larger than 0.1\,$\mu$m have been suggested to explain a number of observations of RSGs and of VY~CMa in particular.
\citet{2009A&A...498..127V} found that an additional continuum opacity source was required to model the spectra of many RSGs. 
Radiative transfer modelling by \citet{2001ApJ...557..844H} of VY~CMa itself suggested that the size distribution of grains extends to several micron, although with an average grain size of only 15\,nm because small dust grains were included.
Meanwhile, \citet{2001AJ....121.1111S} found that the resolved mid-IR emission is consistent with the presence of grains with size $\sim$\,0.3\,$\mu$m.

Our observations provide the first direct confirmation of grains in this size range, approximately 50 times larger than the average size of interstellar medium (ISM) dust \citep{1977ApJ...217..425M}.
As RSGs emit the bulk of their radiation at wavelengths of a few micron, sub-micron dust grains can receive a significant amount of radiation pressure by scattering, rather than absorbing, the stellar emission \citep{2008A&A...491L...1H,2012A&A...546A..76B}.
This has been found to be an effective mechanism for driving mass loss in oxygen-rich AGB stars \citep{2012Natur.484..220N}. 
Further observational and modelling work exploring this possibility in RSGs is required before any definitive conclusions can be made regarding the driving mechanism.

Although both regions examined show evidence for large dust grains, the grains in the South Knot are $\sim$ 50\,\% larger than those in the Southwest Clump, showing that the differences between discrete ejection events extend to the conditions surrounding dust formation and processing.
The two clumps also have substantially different total velocities \citep[][S Knot $\sim$ 42\,km\,s$^{-1}$, SW Clump $\sim$ 18\,km\,s$^{-1}$]{2007AJ....133.2716H}, which raises the question of whether and how this may be related to the differences in grain size.

\subsection{Influence of multiple scattering}
Previous studies \citep{2013AJ....146...90S,2015AJ....150...15S} have suggested that the clumps in the ejecta have optical depths that make the single-scattering approximation inappropriate.
The aforementioned works \citep{2013AJ....146...90S,2015AJ....150...15S} assumed that the ejecta is illuminated by the same spectrum that we observe.
However, given the probable geometry, the regions where we are attempting to measure the grain size may be illuminated by a considerably stronger flux in the wavelength ranges used to measure their optical depth than implied by the observed spectrum.
This means that significantly less material is required to produce the observed scattered flux, and hence the optical depths are lower than previously inferred, improving the chances that the single-scattering approximation is acceptable.

This limit for this approximation is typically set at $\tau_{sca} \sim 0.2$, where Poisson statistics show that 10\% of scattered photons have been multiply scattered.
However, as the alternative is computationally intensive Monte Carlo radiative transfer model fitting, we argue that single scattering is an acceptable solution, provided that the effects of multiple scattering are understood.
In general, multiple scattering introduces depolarisation, reducing the observed degree of polarisation.
In the case of forward scattering (e.g.the South Knot) $p$ typically increases for larger grains, meaning that our size determination is a lower limit.
However, for backward-scattering regions, depolarisation would imply the grains are in fact smaller.

Moreover, given the very high observed polarisation fractions, any depolarising effect cannot be large \citep{2015AJ....150...15S}. 
This further implies that the observed polarisation does not deviate significantly from single scattering, even if there is a significant fraction of multiply-scattered photons.
Therefore we can place robust constraints on the grain size distribution in all cases. 

\subsection{Supernova resistance}

The large grains we have observed are substantially more likely to survive the supernova explosion that will eventually consume VY~CMa.
We created spherically symmetric radiative transfer models using v2.02.70 of {\sc MOCASSIN} \citep{2003MNRAS.340.1136E,2005MNRAS.362.1038E,2008ApJS..175..534E} to determine the sublimation distance of silicate grains of various sizes.
We find that 0.5\,$\mu$m grains would survive sublimation by the supernova radiation\footnote{Assumed to be a 10$^9$ L$_\odot$ source with a blackbody temperature of 10\,000\,K} $\sim 50\%$ closer to the star than ISM grains, which substantially alters the amount of dust to survive the explosion.
The future evolutionary path of VY~CMa is highly uncertain, but if the explosion will not occur for at least another 10\,000\,yr then all of the $\sim 10^{-2}$\,M$_\odot$ of dust formed so far \citep{2001ApJ...557..844H,2007ApJ...656.1109M} will be injected into the ISM given the present outflow velocity.
Although this is small compared to the dust mass in SN\,1987A \citep[$\sim$0.7\,M$_\odot$,][]{2015MNRAS.446.2089W}, it is non-negligible compared to the amounts found in other young ($<$~1000 days) supernova remnants \citep[see][and references therein]{Gallrev}.

\section{Summary}

We have presented the first unambiguous detection of sub-micron dust grains in the outflow of a red supergiant, VY~CMa, based on optical polarimetric imaging with ZIMPOL. 
These grains, with an average size of 0.5\,$\mu$m, are sufficiently large to derive a significant amount of radiative acceleration from scattering, which may indicate that the \citet{2008A&A...491L...1H} model applies to RSGs as well as oxygen--rich AGB stars.
The increased size of the grains also makes them resistant to sublimation in the supernova that will eventually consume VY\,CMa.
This may result in VY~CMa injecting upwards of 10$^{-2}$\,M$_\odot$ of pre-supernova dust into the ISM. 

 \begin{acknowledgements}
We thank the anonymous referee for their helpful suggestions. We thank Josef Hron and Endrik Kr\"ugel for helpful discussions and advice, and . P.S. is supported under DFG programme no. WO 857/10-1. N.V.V. acknowledges support from RFBR grant 13-02-00138a. 
 \end{acknowledgements}

\bibliographystyle{aa}
\bibliography{vycma}

\Online
\begin{appendix}

\section{Grain size fitting}\label{sec_fit}

In order to calculate the maximum likelihood grain-size distribution, we compute the scattering and polarising properties of a grid of dust models under the assumption of Mie scattering \citep{Mie1908} by bare, compact spheres, which allows us to fit the observations under the assumption of single scattering. 
We assume a grain size distribution of the form $n\left(a\right) \propto a^{-q}$, where a is the grain radius and $n\left(a\right)$ is the number density of grains with radius a.
The lower- (a$_{\rm min}$) and upper- (a$_{\rm max}$) size limits are free parameters, and the exponent of the grain size distribution is fixed to $q=3.5$.
For each size distribution, we calculate the scattering and absorbtion cross-sections, and the components of the M\"uller matrix at one degree intervals, for each wavelength for which we wish to fit observations.

The polarisation fraction at each wavelength is completely determined by the elements of the M\"uller matrix, such that $p\left(\lambda, \theta\right) = \left| S_{12}\left(\lambda,\theta\right) / S_{11}\left(\lambda,\theta\right)\right|$, where p is the polarisation fraction, $\lambda$ is the wavelength of interest, $\theta$ is the scattering angle, and S$_{ij}$ are the elements of the M\"uller matrix.

The ratio of the scattered intensities at any pair of wavelengths is determined by the scattering phase function, contained within S$_{11}$, and the ratio of the scattering efficiencies, i.e.

\begin{align}
 \mathcal{R}\left(\theta\right)  =&\, I\left(\lambda_1,\theta\right) / I\left(\lambda_2,\theta\right)\nonumber\\ 
  =& \left(Q_{\rm sca}\left(\lambda_1\right) / Q_{\rm sca}\left(\lambda_2\right) \right) \times \left( S_{11}\left(\lambda_1,\theta\right) / S_{11}\left(\lambda_2,\theta\right) \right)\\ &\quad\times \left( F_\ast\left(\lambda_1\right) / F_\ast\left(\lambda_2\right) \right),\nonumber
\end{align}
where $F_\ast$ is the flux emitted by the star and $Q_{\rm sca}$ are the scattering efficiencies.

For each grain-size distribution, we calculate the likelihood
$$\mathcal{L}\left(\Theta\right) = \prod L_i\left(\Theta,\Delta_i\right) $$

$$\mathcal{L}\left(\Theta\right)= \prod \frac{\sigma_{\Delta_i}}{2\pi} \exp\left(\frac{-1}{2\sigma^2_{\Delta,i}}\left( \Delta_i - M_i\left(\Theta\right)\right)^2 \right)$$
where $\Delta$ is a vector of observations such that $\Delta_i$ is the {\it i}th observation and $\sigma_{\Delta_i}$ its associated uncertainty, M$_i$ is the output of the model associated with the {\it i}th observation, and $\Theta$ is a vector of model parameters (in our case $\Theta = \left(a_{\rm min}, a_{\rm max}\right)$).
The maximum-likelihood model is then defined to be the model with parameters $\Theta$ that maximises the value of $\mathcal{L}\left(\Theta\right)$, and $\Theta$ contains the information in which we are interested.

For the sake of efficiency, it is common to compute the log-likelihood
$$Log\left(\mathcal{L}\left(\Theta\right)\right)= \sum  \frac{-1}{2\sigma^2_{\Delta,i}}\left( \Delta_i - M_i\left(\Theta\right)\right)^2$$
by taking only the exponent.

\begin{figure}
    	\resizebox{\hsize}{!}{\includegraphics{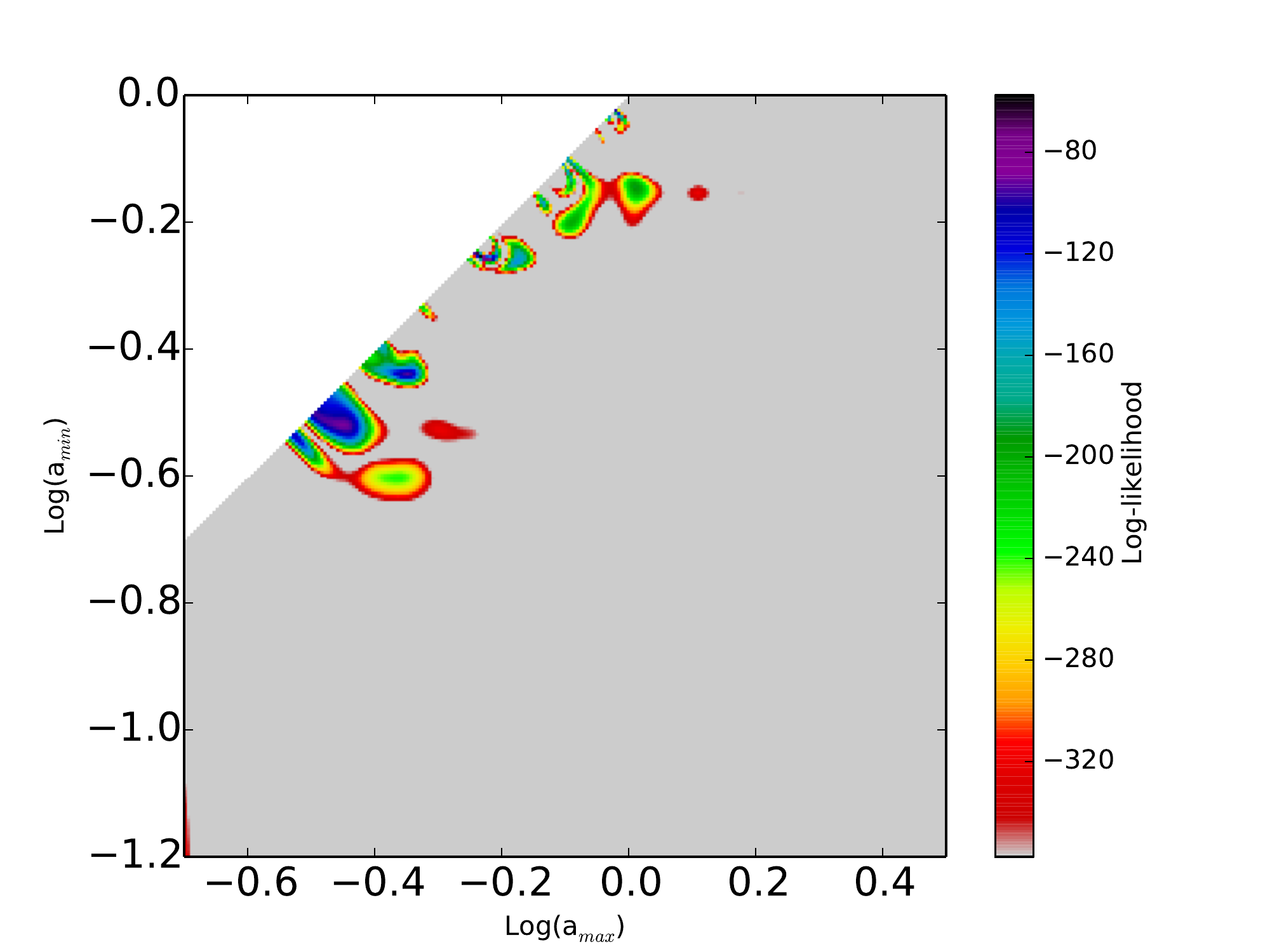}}
\caption{Likelihood space for scattering angle of 63$^\circ$, corresponding to the S Knot, zoomed to show grain sizes in the range 0.1\,--\,3\,$\mu$m.}
\label{fig_like2}
\end{figure}

By calculating a grid of grain-size distributions with 5\,nm $<$ a$_{\rm min} \leq 1 \mu$m and 100\,nm $<$ a$_{\rm max} \leq 50 \mu$m we cover the full range of grain sizes observed in the interstellar medium and in the circumstellar environments of evolved stars. 
The resulting likelihood space for the S Knot is shown in Fig.~\ref{fig_like2}, zoomed to show the region of interest.

The average size of the maximum-likelihood distribution is calculated simply from $$\left\langle a\right\rangle=\int ^{a_{max}}_{a_{min}} a \,\/ n\left(a\right)da\biggm/\int^{a_{max}}_{a_{min}} n\left(a\right)da$$

$$\left\langle a\right\rangle=\int^{a_{max}}_{a_{min}} a^{-q+1}da \biggm/\int^{a_{max}}_{a_{min}} a^{-q}da,$$
which can be solved analytically.

Because the scattering properties are determined by the real part of the complex refractive index of the grain material, which is similar for all of the plausible choices, the optical properties are dominated by the grain size for $a\sim\lambda$, rather than the choice of silicate material.

\end{appendix}

\end{document}